\documentclass[aps,prc,twocolumn,times,graphicx,tighten,showpacs]{revtex4}
\usepackage[dvips]{graphicx}
\usepackage[dvips]{color}
\usepackage{amsmath}

\newcommand{\bea}{\begin{eqnarray}}
\newcommand{\eea}{\end{eqnarray}}
\newcommand{\be}{\begin{equation}}
\newcommand{\ee}{\end{equation}}
\newcommand{\np}{{\bf p}}

\newcommand{\nh}{{\bf h}}

\newcommand{\sumint}{\sum\kern -3.5ex \int\kern 1.0ex}

\newlength\dlf  % Define a new measure, dlf

\begin{document}

%------------------------------------------
\title{Emission of 
neutron-proton and proton-proton pairs in electron
scattering induced by meson-exchange currents 
}

%--------------------
\author{
I. Ruiz Simo$^a$,
J.E. Amaro$^a$,
M.B. Barbaro$^{b,c}$,
A. De Pace$^b$,
J.A. Caballero$^d$,
G.D. Megias$^d$, 
T.W. Donnelly$^e$
}
%--------------------

\affiliation{$^a$Departamento de F\'{\i}sica At\'omica, Molecular y Nuclear,
and Instituto de F\'{\i}sica Te\'orica y Computacional Carlos I,
Universidad de Granada, Granada 18071, Spain}

\affiliation{$^b$INFN, Sezione di Torino, Via P. Giuria 1, 10125 Torino, Italy}
  
\affiliation{$^c$Dipartimento di Fisica, Universit\`a di Torino 
 Via P. Giuria 1, 10125 Torino, Italy}

\affiliation{$^d$Departamento de F\'{\i}sica At\'omica, Molecular y Nuclear,
Universidad de Sevilla, Apdo.1065, 41080 Sevilla, Spain}

\affiliation{$^e$Center for Theoretical Physics, Laboratory for Nuclear
  Science and Department of Physics, Massachusetts Institute of Technology,
  Cambridge, MA 02139, USA}

\date{\today}

%----------------------------------------------------------------------

\begin{abstract}

We use a relativistic model of meson-exchange currents to compute the
proton-neutron and proton-proton yields in $(e,e')$ scattering from
$^{12}$C in the 2p-2h channel. We compute the response functions and
cross section with the relativistic Fermi gas model for a range of
kinematics from intermediate to high momentum transfers.  We find a
large contribution of neutron-proton configurations in the initial
state, as compared to proton-proton pairs.  The different emission
probabilities of distinct species of nucleon pairs are produced in our
model only by meson-exchange currents, mainly by the $\Delta$ isobar
current. We also analyze the effect of the exchange contribution and
show that the direct/exchange interference strongly affects the
determination of the np/pp ratio.

\end{abstract}

%-----------------------------------
\pacs{25.30.Fj; 21.60.Cs; 24.10.Jv}
%-------------------------------------

\maketitle

The process of two-nucleon knockout reactions in electron scattering
is thought to involve short-range correlations (SRC) in nuclei.  In
this case one expects an excess of neutron-proton (np) pairs over
proton-proton (pp) pairs. Experiments on $^{12}$C for high momentum
transfer and missing momentum, have reported a number of np pairs
18$\pm$5 times larger than their pp counterparts
\cite{Shn07,Sub08,Ryc15}.  More recently the dependence on the nuclear
mass number has been studied in \cite{Hen13,Hen14,Col15}, where the
aim of this series of recent investigations at Jefferson Lab is to
unambiguously determine the short-range properties of light nuclei
\cite{Fom12,Mon14,Kor14,Egi06}. The analysis of these experiments
seems to be in agreement with theoretical predictions of nucleon and
nucleon pair momentum distributions in variational Monte Carlo
calculations, where the importance of the tensor forces in the
ground-state correlations of nuclei has been emphasized
\cite{Sch07,Wir14}.  Note that in the mentioned electron scattering
experiments the kinematics involved high momentum transfers, $Q^2>
1.4$ (GeV/c)$^2$, and thus relativistic corrections are likely to be
important in the theoretical descriptions employed.

Another source of SRC evidence comes from calculations of the
semi-inclusive electron scattering reaction $(e,e'pN)$, that however
rely on factorization approximations that have not been fully
justified for all the kinematics of interest \cite{Col16}.  While the
kinematics of the experiments have been selected to minimize the
contribution from other mechanisms that can induce two-particle
emission, such as meson-exchange currents (MEC) and isobar excitations
\cite{Shn07}, the contribution of MEC cannot be ruled out {\em a
  priori}.

Similarly to the electron case, observation of events in neutrino
scattering with a pair of energetic protons has been reported in the
ArgoNeuT experiment \cite{Acc14}. From these events several
back-to-back nucleon configurations have been identified and
associated to nuclear mechanisms involving short-range correlated
neutron-proton (np) pairs in the nucleus \cite{Cav15}.  The SRC
explanation of this excess of back-to-back events in ArgoNeuT is still
controversial \cite{Wei16,Nie16}.

In this work we investigate the relative effects of MEC on the
separate pp and np channels in the inclusive 2p-2h cross section,
without including NN correlations. It is important to know if the MEC
alone can explain, at least partially, the observed enhancement of the
$^{12}$C$(e,e'np)$ cross section over that of the $^{12}$C$(e,e'pp)$
cross section, as observed in the data \cite{Sub08}. This is in lieu
of a fully reliable relativistic model for the $(e,e'pN)$ cross
section, as this is unavailable.

 We have recently developed a fully relativistic model of
 meson-exchange currents in the 2p-2h channel for electron and
 neutrino scattering \cite{Sim16}. This model is an extension of the
 relativistic MEC  model of \cite{Pac03} to the weak
 sector. It has been recently validated by comparing to the
 $^{12}$C$(e,e')$ inclusive cross section data for a wide kinematical
 range within the SuperScaling approach (SuSA) \cite{Meg16}. This
 model describes jointly the quasielastic and inelastic regions using
 two scaling functions fitted to reproduce the data, while the 2p-2h
 MEC contribution properly fills the dip region in between, resulting
 in a excellent global agreement with the data.

With this benchmark model we are able to study the separate np and pp
channels in the response functions and cross section for the three
$(e,e')$, $(\nu_l,l^-)$ and $(\bar{\nu}_l,l^+)$ reactions. Although
our 2p-2h model does not explicitly include NN correlations, they are
implicitly accounted for in the phenomenological scaling functions,
and thus we cannot isolate the 2p-2h contributions coming from SRC in
this approach. However, with this model we are at least able to
provide a precise estimation of the size of MEC in the separate
channels for high momentum and energy transfers where relativistic
effects are important. In this work we focus on the contributions of
pion-in-flight, seagull, and $\Delta$(1232) excitation diagrams of the
MEC.

\begin{figure}
\begin{center}
\includegraphics[scale=0.7,bb=160 430 400 790]{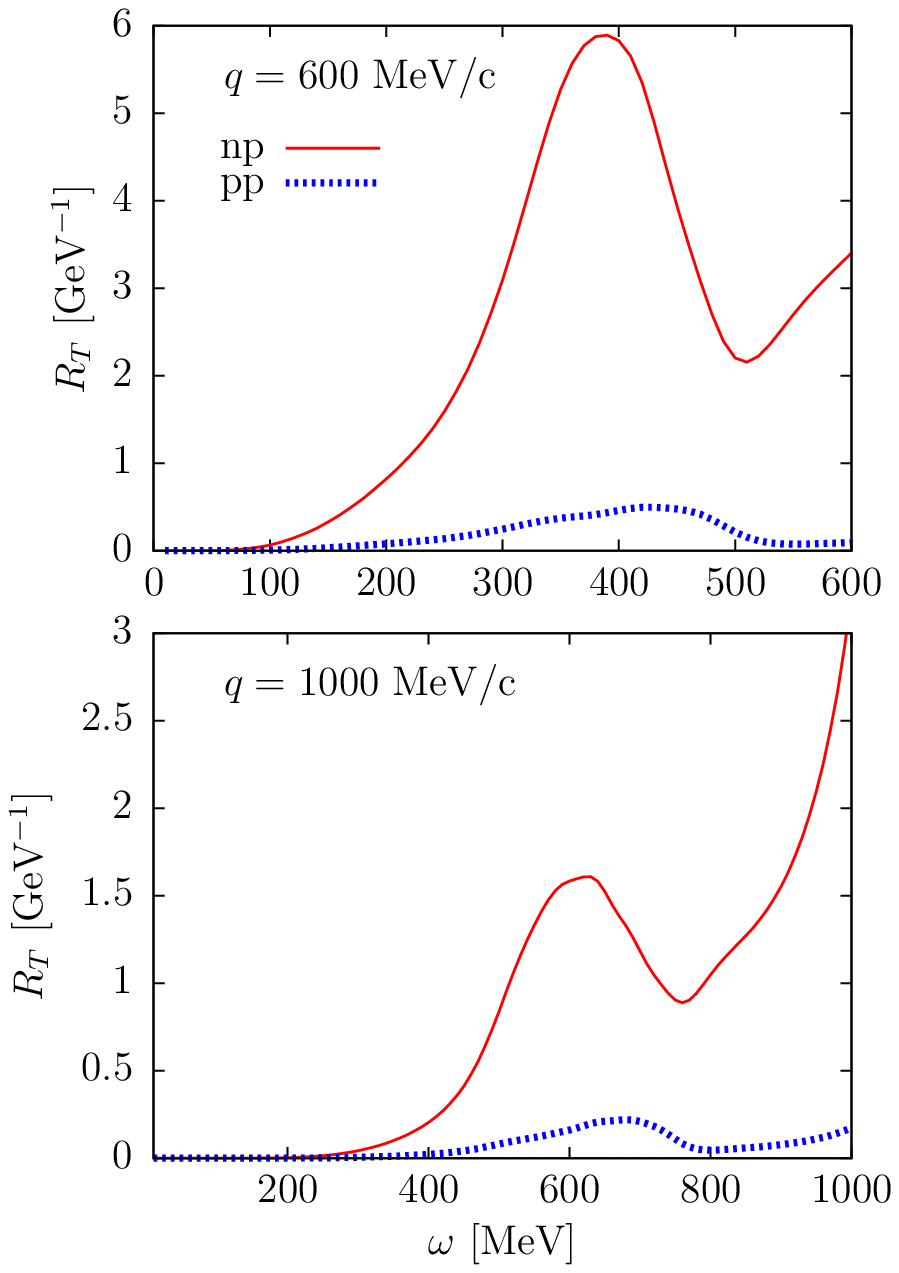}
\caption{(color online) Electromagnetic transverse response function for np
  and pp pair emission off $^{12}$C as a function of $\omega$ for two values
  of $q$.  }
\label{fig1}
\end{center}
\end{figure}

\begin{figure}
\begin{center}
\includegraphics[scale=0.7,bb=160 430 400 790]{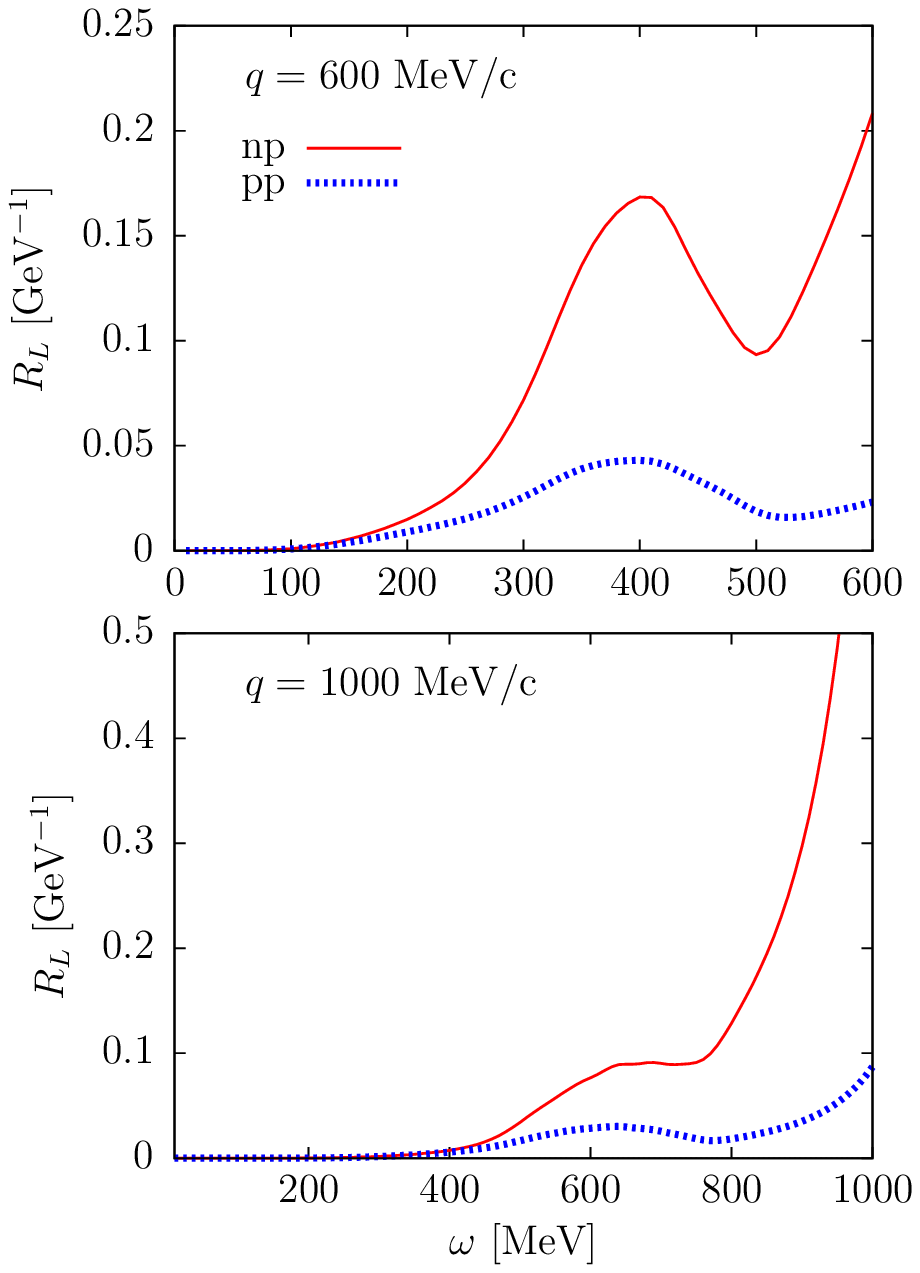}
\caption{(color online) As for Fig. 1, but now for the longitudinal response function.
}
\label{fig2}
\end{center}
\end{figure}

We write the inclusive $(e,e')$ cross section as the product of
the Mott cross section and a linear combination of longitudinal and
transverse response functions:
\begin{equation}
\frac{d\sigma}{d\Omega d\omega}
= \sigma_{\rm Mott}
\left[ v_L R_L(q,\omega)+v_T R_T(q,\omega)\right].
\label{sigma}
\end{equation}
The response functions for momentum transfer $q$ and energy transfer
$\omega$ contain the contribution of all of the nuclear excitations with
energy $\omega$. 

In the relativistic Fermi gas model the excitations can be 1p-1h,
2p-2h and so on.  We describe the particle and hole states as
relativistic plane waves with momenta above and below the Fermi
momentum $k_F$, respectively. 
 In this work we compute the 2p-2h
contributions to the response functions that are
 proportional to the volume $V$ of the system, 
which for symmetric nuclear matter has $Z=N=A/2$, where  
 $V=3\pi^2 Z/k_F^3$.  They are given by
\begin{eqnarray}
R^K_{2p-2h}
&=&
\frac{V}{(2\pi)^9}\int
d^3p'_1
d^3h_1
d^3h_2
\frac{m_N^4}{E_1E_2E'_1E'_2}
\nonumber \\ 
&\times&
r^{K}(\np'_1,\np'_2,\nh_1,\nh_2)
\delta(E'_1+E'_2-E_1-E_2-\omega)
\nonumber\\
&\times&
\theta(p'_2-k_F)
\theta(p'_1-k_F)
\nonumber\\
&\times& 
\theta(k_F-h_1)
\theta(k_F-h_2),
\label{hadronic}
\end{eqnarray}
for $K=L,T$, where $\bf p'_2= h_1+h_2+q-p'_1$ is fixed by momentum
conservation, $m_N$ is the nucleon mass, $E_i$ and $E'_i$ are the
on-shell energies of the holes and particles, respectively. 
The response functions for the elementary 2p-2h excitation,
$r^{K}(\np'_1,\np'_2,\nh_1,\nh_2)$, for given initial and final
momenta, are the sums over spins of the squares of the MEC matrix
elements and can be found in \cite{Sim16} for the separate np, pp and
nn charge channels.  The 2p-2h states are antisymmetrized and
therefore our MEC matrix elements and response functions contain
direct and exchange contributions. The 7D integral of
Eq. (\ref{hadronic}) is computed numerically without approximations by
following the method of \cite{Ruiz14,Ruiz14b}.  We refer the reader to
\cite{Sim16} for further details on the model.

\begin{figure}
\begin{center}
\includegraphics[scale=0.5,bb=80 470 480 790]{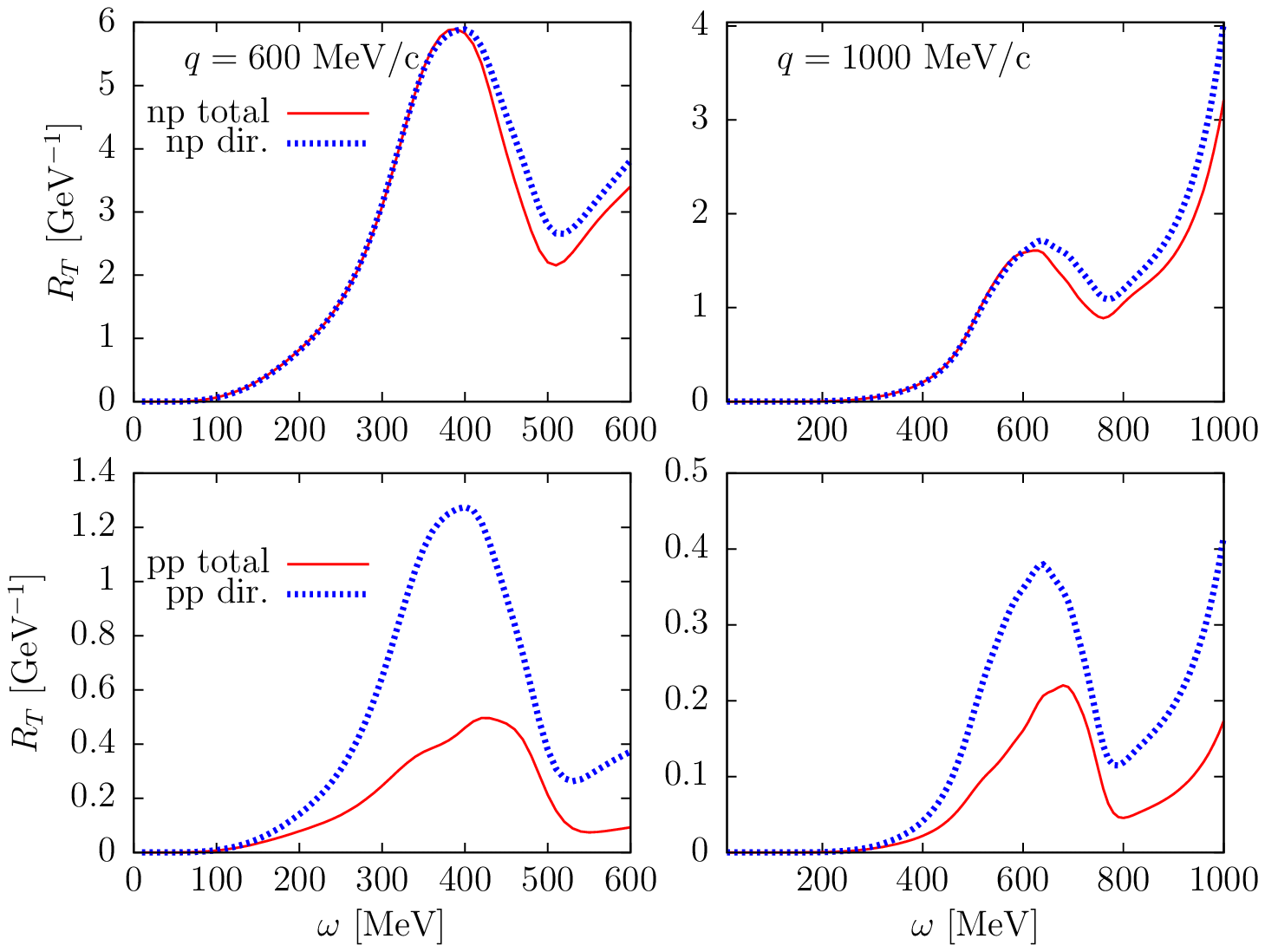}
\caption{(color online) Electromagnetic transverse response function showing the contributions of 
direct-exchange interference terms to the separate np and pp channels.
}
\label{fig3}
\end{center}
\end{figure}

\begin{figure}
\begin{center}
\includegraphics[scale=0.77,bb=180 420 430 790]{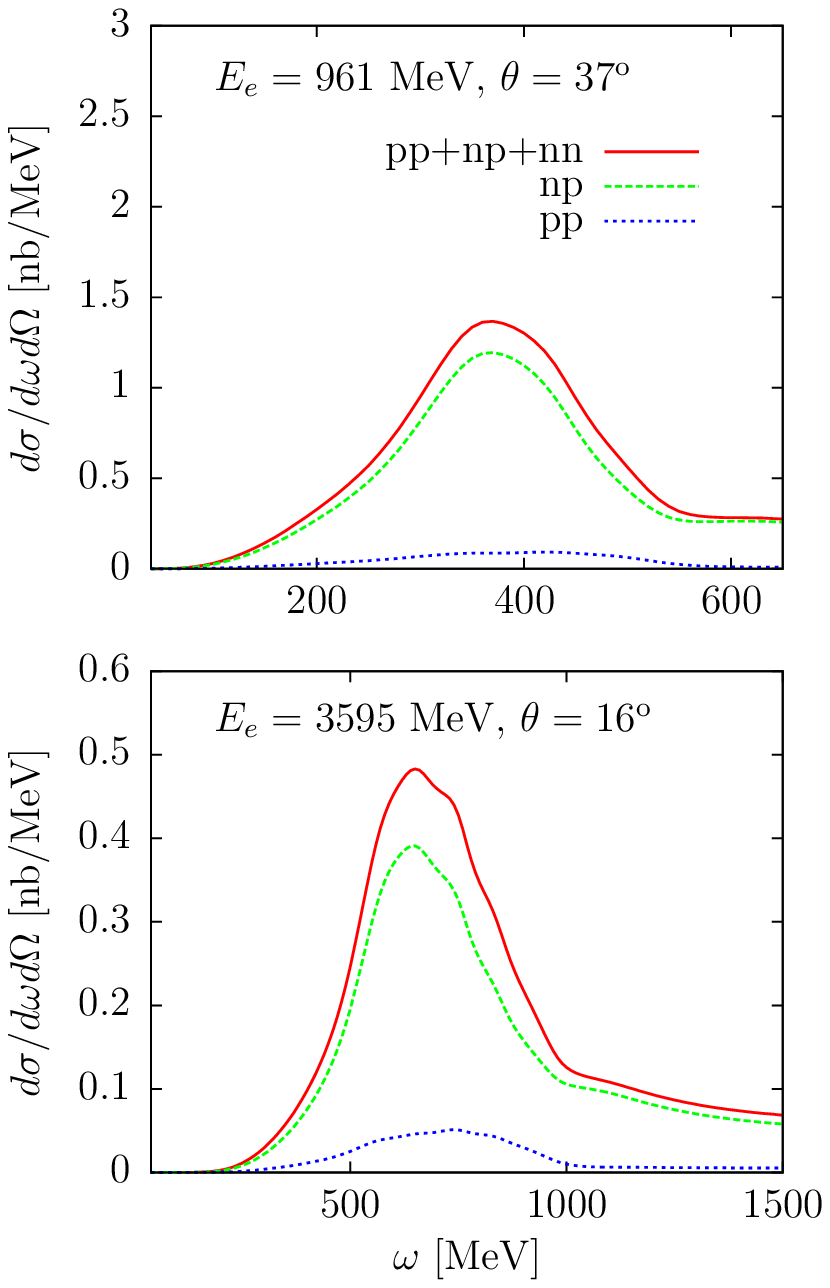}
\caption{(color online) Electromagnetic differential cross section separated into the different
charge contributions. 
}
\label{fig4}
\end{center}
\end{figure}

\begin{figure}
\begin{center}
\includegraphics[scale=0.77,bb=180 420 430 790]{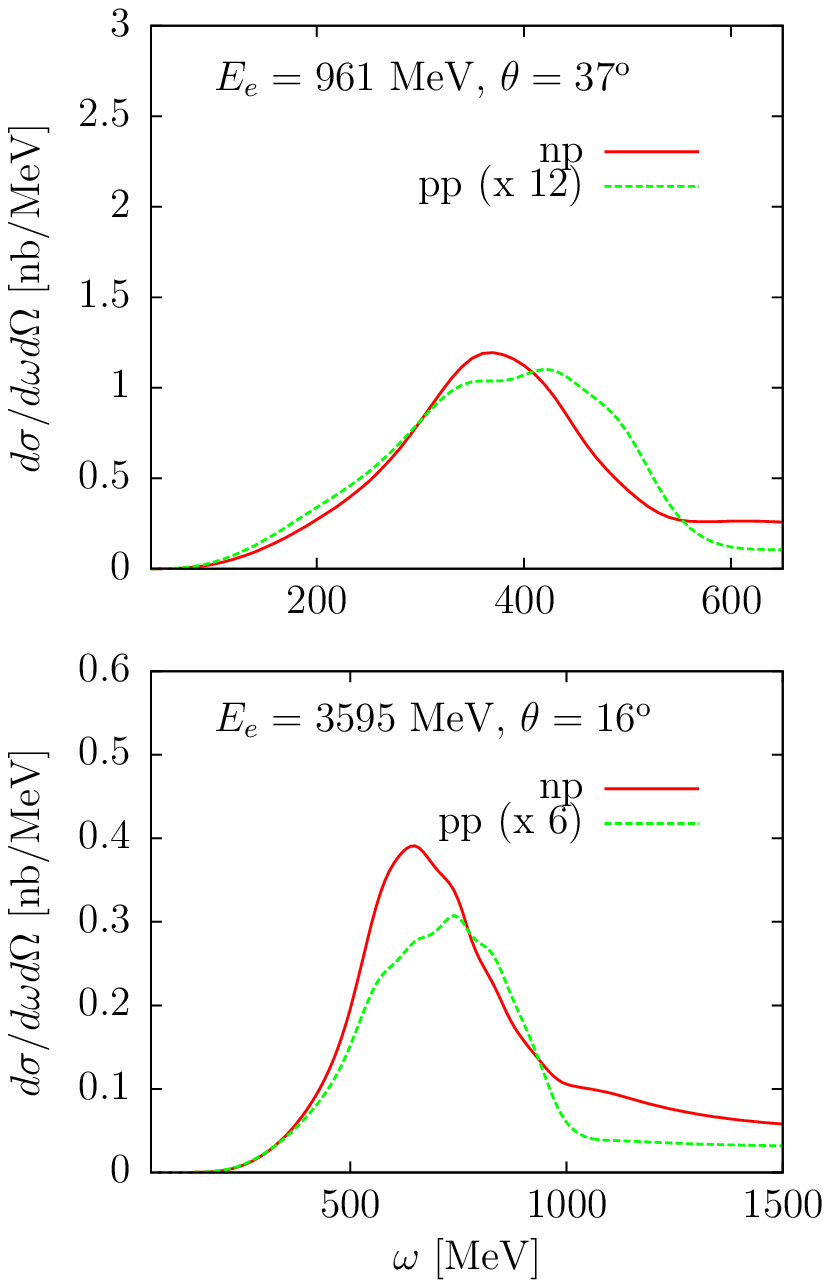}
\caption{(color online) As for Fig. 4, but now the pp contribution has been scaled
by a constant factor, as shown in the two panels.  
}
\label{fig5}
\end{center}
\end{figure}

\begin{figure}
\begin{center}
\includegraphics[scale=0.7,bb=180 620 430 790]{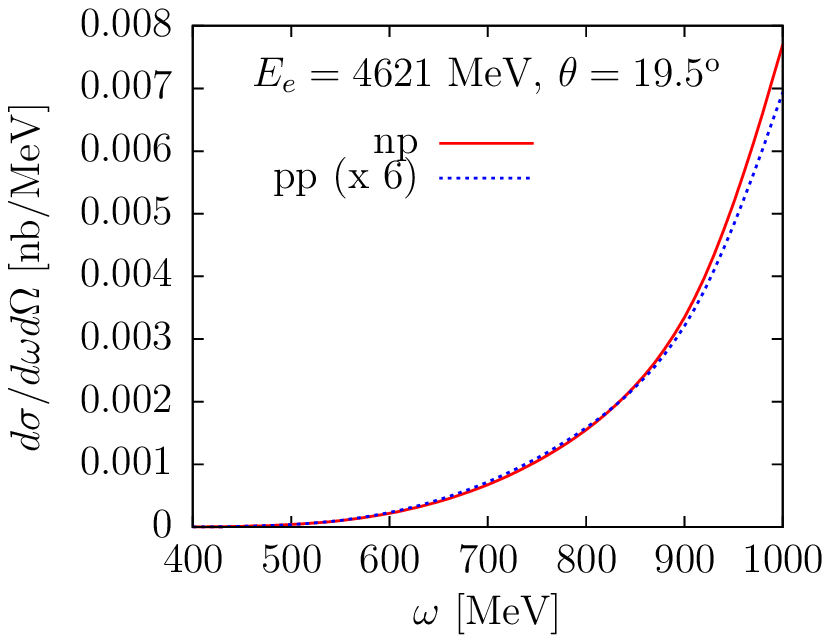}
\caption{(color online) As for Fig. 5, but now for the electron kinematics of the experimental
setup of \cite{Shn07}.
}
\label{fig6}
\end{center}
\end{figure}

In what follows we show results for the separate np and pp pair
emission from $^{12}$C. In our model, the nn channel gives the same
contribution as the pp one, because both are induced by the
$\Delta$-isobar current.  In Figs. 1, 2 we show the transverse and
longitudinal response functions. For $q=600$ MeV/c, the np transverse
response is around a factor of twelve times larger than the pp
one. For $q=1000$ MeV/c this factor gets reduced to $\sim 6$. In the
the longitudinal responses the np/pp ratio is further
reduced. However, the longitudinal MEC contribution to the cross
section is almost negligible compared with the transverse one, because
the dominant $\Delta$ excitation current is mainly transverse.

In Fig. 3 we show the effect of neglecting the direct-exchange
interference of the MEC matrix elements. For np pair emission it is
negligible. On the other hand, for pp emission it is of the same order
as the direct contribution.  Therefore neglecting this interference
would reduce the above mentioned np/pp ratios by a factor of two.
This implies that the interference is crucial for describing this
ratio properly. Although the net effect of the interference in the
2p-2h is less than 20\% and sometimes it can be safely disregarded
\cite{Car94,Gil97}, this is not the case for the pp separate
contribution. However, up to now there have been no calculations of these
ratios.

In Fig. 4 we show the semi-inclusive double-differential cross section
for pair emission from $^{12}$C for two incident energies and electron
scattering angles. The separate np and pp pair emission channels are
also displayed. As in the case of the transverse response, the np pair
emission clearly dominates over pp. The ratio of np over pp depends on
the kinematics. It is roughly between twelve and six, as can be seen
in Fig. 5, where we have multiplied the pp cross section by a constant
factor. It is worth mentioning that the np/pp ratio also depends on
$\omega$, although a constant scaling factor does a reasonable job at 
inter-relating the different channels.

In Fig. 6 we show the separate np and pp emission cross sections for
the kinematics of the measurement of \cite{Shn07,Sub08}. In this
experiment the energy transfer was chosen $\omega=865$ MeV, well below
the $\Delta$ excitation maximum which occurs at about 1.12 GeV.  For
this reason we only show in Fig. 6 the low-energy tail of the cross
section below this value of $\omega$. These kinematics were chosen to
minimize the MEC contribution that we are showing here. In fact, for
this value of $\omega$ the MEC are small, since we are far from the
maximum of the $\Delta$ peak.  The results of Fig. 6 are the expected
MEC contributions to the semi-inclusive $(e,e'pN)$ cross section in an
uncorrelated system. In this case the np/pp ratio is a factor 6. This
is not sufficient to explain the factor 18$\pm$5 found in the
experiment and attributed to SRC, coming mainly from the tensor
nuclear force.  
However, it does suggest that any analysis
where MEC effects are not included must be viewed with
caution. 
In the MEC
case this factor comes roughly from isospin considerations, but the
effect of the direct-exchange interference is $q$ dependent and can
modify it by a large extent depending on the kinematics. Note that the
kinematics of the experiment \cite{Shn07} is relativistic, with
$Q^2=2$ (GeV/c)$^2$, and $q=1.65$ GeV/c. Thus the relativistic
calculation of the 2p-2h MEC responses is mandatory.  One of the
findings of \cite{Shn07,Sub08} was that there are very few correlated
pp pairs. Therefore we expect that our result for the pp cross section
will not change significantly when including SRC, while the np cross
section should be considerably increased.

In summary, we have computed the semi-inclusive $^{12}$C$(e,e'np)$ and
$^{12}$C$(e,e'pp)$ cross sections in the RFG including a fully
relativistic model of MEC.  The np/pp ratio has been quantified and
analyzed for several kinematics. The direct-exchange interference
terms are found to be important, especially in the pp channel.  The
MEC alone are not able to explain completely the data for this ratio,
which is found to be larger than our findings, although they clearly
should be expected to play an important role in determining the np/pp
ratio in two-nucleon emission electron scattering, as well as related
flavor dependence in charge-changing neutrino reactions. Said another
way: while missing contributions in our model (like SRC) could be
important at least for the kinematics of existing experiment, our
results indicate that, in order to understand in depth the size of
such effects, the MEC contributions should also be included. In the
future the relativistic modeling used for the latter could be extended
to include correlation currents of the pionic type \cite{Ama10}, effective
interactions or, alternatively, correlation operators in the wave
functions.

%%%%%%%%%%%%%%%%%%%%%%%%%%%%%%%%%%%%%%%%%%%%%%%%%%%%%%%%%%%%%%%%%%%%%
\section*{Acknowledgments}
This work was supported by Spanish Direcci\'on General de
Investigaci\'on Cient\'ifica y T\'ecnica and FEDER funds (grants
No. FIS2014-59386-P and No. FIS2014-53448-C2-1), by the Agencia de
Innovaci\'on y Desarrollo de Andaluc\'ia (grants No. FQM225, FQM160),
by INFN under project MANYBODY, and part (TWD) by U.S. Department of
Energy under cooperative agreement DE-FC02-94ER40818. IRS acknowledges
support from a Juan de la Cierva-incorporaci\'on fellowship from
Spanish MINECO.  GDM acknowledges support from a Junta de Andalucia fellowship
(FQM7632, Proyectos de Excelencia 2011).
The authors also acknowledge support from ``Espace de Structure et de
r\'eactions Nucl\'eaire Th\'eorique" (ESNT, http://esnt.cea.fr ) at
CEA-Saclay, where this work was partially carried out.
%%%%%%%%%%%%%%%%%%%%%%%%%%%%%%%%%%%%%%%%%%%%%%%%%%%%%%%%%%%%%%%%%%%%

\end{document}